\documentclass[onecolumn,aps,showpacs,superscriptaddress,pre,preprint]{revtex4}
\usepackage{hyperref}
\usepackage{graphicx}

\begin{document}

\title{Anti-persistent behavior of defects in a lyotropic liquid crystal during annihilation}

\author{H. V. Ribeiro}\email{hvr@dfi.uem.br}
\affiliation{Departamento de F\'isica, Universidade Estadual de Maring\'a, 87020-900, Maring\'a, Paran\'a, Brazil}
\affiliation{National Institute of Science and Technology for Complex Systems, CNPq, 22290-180, Rio de Janeiro, Rio de Janeiro, Brazil}
\author{R. R. Guimar\~aes}
\affiliation{Departamento de F\'isica, Universidade Estadual de Maring\'a, 87020-900, Maring\'a, Paran\'a, Brazil}
\affiliation{National Institute of Science and Technology for Complex Systems, CNPq, 22290-180, Rio de Janeiro, Rio de Janeiro, Brazil}
\affiliation{National Institute of Science and Technology for Complex Fluids, CNPq, 05508-090, S\~ao Paulo, S\~ao Paulo, Brazil}
\author{R. T. Teixeira-Souza}
\affiliation{Universidade Tecnol\'ogica Federal do Paran\'a, Campus Pato Branco, 85503-390, Pato Branco, Paran\'a, Brazil}
\author{\mbox{H. Mukai}}
\author{P. R. G. Fernandes}
\affiliation{Departamento de F\'isica, Universidade Estadual de Maring\'a, 87020-900, Maring\'a, Paran\'a, Brazil}
\affiliation{National Institute of Science and Technology for Complex Fluids, CNPq, 05508-090, S\~ao Paulo, S\~ao Paulo, Brazil}
\author{E. K. Lenzi}
\affiliation{Departamento de F\'isica, Universidade Estadual de Maring\'a, 87020-900, Maring\'a, Paran\'a, Brazil}
\affiliation{National Institute of Science and Technology for Complex Systems, CNPq, 22290-180, Rio de Janeiro, Rio de Janeiro, Brazil}
\author{R. S. Mendes}
\affiliation{Departamento de F\'isica, Universidade Estadual de Maring\'a, 87020-900, Maring\'a, Paran\'a, Brazil}
\affiliation{National Institute of Science and Technology for Complex Systems, CNPq, 22290-180, Rio de Janeiro, Rio de Janeiro, Brazil}

\date{\today}

\begin{abstract}
We report on the dynamical behavior of defects of strength $s = \pm\,1/2$ in a lyotropic liquid crystal during the annihilation process. 
By following their positions using time resolved polarizing microscopy technique, we present statistically significant evidence that the relative 
velocity between defect pairs is Gaussian distributed, anti-persistent and long-range correlated. We further show that simulations 
of the Lebwohl-Lasher model reproduce quite well our experimental findings.
\end{abstract}

%61.30.Jf	Defects in liquid crystals
%61.30.St	Lyotropic phases
%05.45.Tp	Time series analysis
\maketitle

\textbf{\textit{Introduction -}} The understanding of ordering processes in condensed matter has been the focus of considerable research over the last decades~\cite{chuang,turok,finn,yurke, toyoki, zapotocky, marshall, dutta, rey, oliveira, oliveira2}. One of the main aspects of the ordering process is the dynamical behavior of defects present in the material. Topological defects appear in several systems that present some kind of ordering~\cite{kibble,charlier,figueiras,petit,abu,carvalho}.  
Alloys~\cite{hamad}, semiconductors~\cite{emtsev}, polymers~\cite{quarti} and liquid crystals~\cite{pasini,mukai} are just a few examples. 
In particular, optical textures of liquid crystals~\cite{chandra,degennes,figueiredo_book} are known to be an excellent system for studying the dynamical behavior of defects.
In fact, there are several studies on the annihilation dynamics of defects in liquid crystals employing thermotropics~\cite{marshall,cypt,mendez,minoura},
thermotropic polyester materials~\cite{shiwaku,ding,wang} and lyotropic~\cite{renato}. Moreover, the experimental 
analysis of annihilation processes has also motivated many numerical simulations~\cite{yurke,zapotocky,oliveira,svensek,svetec}. One of the most striking
results of these studies is the scaling law in the relative distance $D(t)$ between defect and anti-defect during the annihilation process, that is,
$D(t) \sim t^\alpha$, where $t$ is the remaining time for the annihilation and $\alpha$ is the power law exponent. It is also known that defect and 
anti-defect present an anisotropic behavior when approaching each other, the defect usually moves faster than the anti-defect~\cite{toth, toth2, svensek2, oswald,blanc}. This last aspect has been recently addressed both, experimentally and numerically, by Dierking~\textit{et al.}~\cite{dierking2},
for umbilical defects in a thermotropic liquid crystal under applied electric field. 

It is surprising that almost no attention has been paid to understand higher order properties of defect trajectories. In this brief report,
we fill some of these lacunas by investigating correlational aspects and the velocity distribution of the defects. Specially,
we present statically significant evidence that the relative velocity between defects pairs in a lyotropic liquid crystal is Gaussian 
distributed, anti-persistent and long-range correlated. Furthermore, we show that these behaviors also appear in our numerical simulations of 
the Lebwohl-Lasher model~\cite{leblas} by using a Langevin approach~\cite{yurke,bac}. In following, we present the experimental setup used to obtain the liquid crystal textures, the image technique employed
to follow the defect motions, the analysis of the experimental data, the Lebwhol-Lasher model used for reproducing our experimental findings and, finally,
a summary of our results.

\begin{figure}[!ht]
\centering
\includegraphics[scale=0.50]{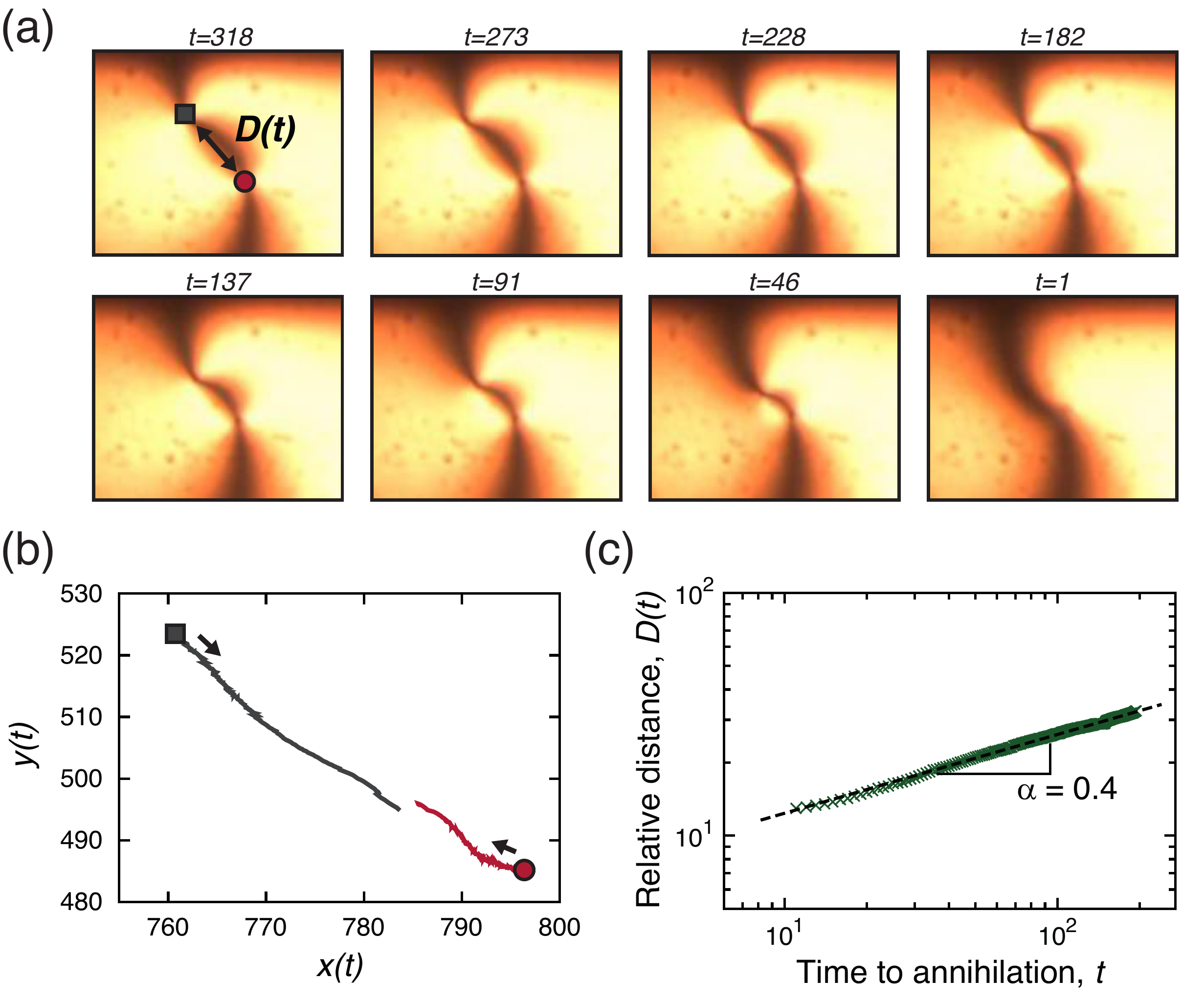}
\caption{(Color online) Annihilation process of defects in our lyotropic system. (a) Typical evolution of textures during the annihilation. The $t$
value (above of each image) stands for the remaining time (in minutes) to the annihilation. The square represents the initial position of the defect and the circle is the same for the anti-defect. The arrow indicates the relative distance between the defect and the anti-defect. (b) The trajectories of the two defects obtained via the Lucas-Kanade algorithm~\cite{Lucas}. The distance units are given in pixels. (c) The scaling law describing the relative distance between the defects,
that is, $D(t) \propto t^{0.4}$.}
\label{fig1}
\end{figure}

\textbf{\textit{Experimental Setup -}} The lyotropic system we have studied is a mixture of potassium laurate (KL~$\rightarrow$~27.49~wt\%), decanol (DeOH~$\rightarrow$ 6.24 wt\%) and distilled and deionized water (H$_2$O~$\rightarrow$~66.27~wt\%). 
At these concentrations, our lyotropic system presents an isotropic $\to$ nematic transition
at $\approx40^\circ$ Celsius. Furthermore, Mukai~\textit{et al.}~\cite{mukai} have observed the defects in lyotropic systems 
are more stable than those ones obtained in thermotropic liquid crystals.
We have made a series of 8 samples of this lyotropic system by using sealed
films in flat capillaries (100~$\mu$m thick by 2~mm width by 2~cm length). These samples were placed in a polarized light microscope coupled with a hostage (INSTEC model HCS302), enabling the precise control of the temperature with ${0.01}^\circ$ Celsius accuracy. In order to produce the defects, we have promoted a spontaneous symmetry breaking~\cite{mukai,kleman} by raising the temperature of the sample to $50^\circ$ Celsius (isotropic region) and suddenly changing it to $25^\circ$ Celsius (nematic region). After this temperature change, we first observed colorful domains, and next ($\approx 5$ hours later) the defects become
visible. The defects observed in these samples are characterized by a strength $s = \pm1/2$, since they have two dark branches from a dark point.
We thus started to capture snapshots of the evolution of so called ``Schlieren'' texture~\cite{dierking_book,figueiredo_book} at a sample rate of 15 pictures by hour until the vanishing of all defects in the sample ($\approx20$ hours after the phase transition). 
Figure~\ref{fig1}(a) shows characteristic textures where the dark branches correspond to areas where the director $\vec{n}$ is perpendicular or parallel to one of the polarizers, while in the bright regions represent areas where the director $\vec{n}$ is tilted compared with the polarizers. The defect is exactly located between the junction of two dark branches, where the orientation of director $\vec{n}$ is not defined. In general, there appear about 20 pairs of defects across the sample and usually 5 of them fall within the microscope region of view.

\textbf{\textit{Data analysis -}} By using sequences of textures of the annihilation process, we track the motion of the defect pairs using the Lucas-Kanade algorithm~\cite{Lucas}. 
In this algorithm, we give the initial positions of a defect pair in the first image and, it automatically assigns the position of the defects in the
subsequent images. Figure~\ref{fig1}(b) shows an example of trajectories of a defect (upper curve) and an anti-defect (bottom curve). We observe that
the defect moves about twice the distance of the anti-defect, similarly to the previous-mentioned behavior of umbilical defects in a 
thermotropic liquid crystal~\cite{dierking2}. We have tracked the evolution of defects in 49 annihilations.
For sake of definition, let \mbox{$\vec{r}_{+}(t)=(x_{+}(t),y_{+}(t))$} be the position vector of the defect
and \mbox{$\vec{r}_{-}(t)=(x_{-}(t),y_{-}(t))$} represents the same for the anti-defect. Here, we focus our analysis on the relative distance in
a pair of defects $D(t)=\|\vec{r}_{+}(t)-\vec{r}_{-}(t)\|$ and also on the relative velocity $V(t)=\frac{D(t+dt)-D(t)}{dt}$. Figure~\ref{fig1}(c) shows
the scaling law that characterizes the evolution of the relative distance, that is, $D(t) \sim t^\alpha$, with $\alpha = 0.4$ in this case 
(see Ref.~\cite{renato} for a more detailed study of this scaling law).

\begin{figure}[!ht]
\centering
\includegraphics[scale=0.50]{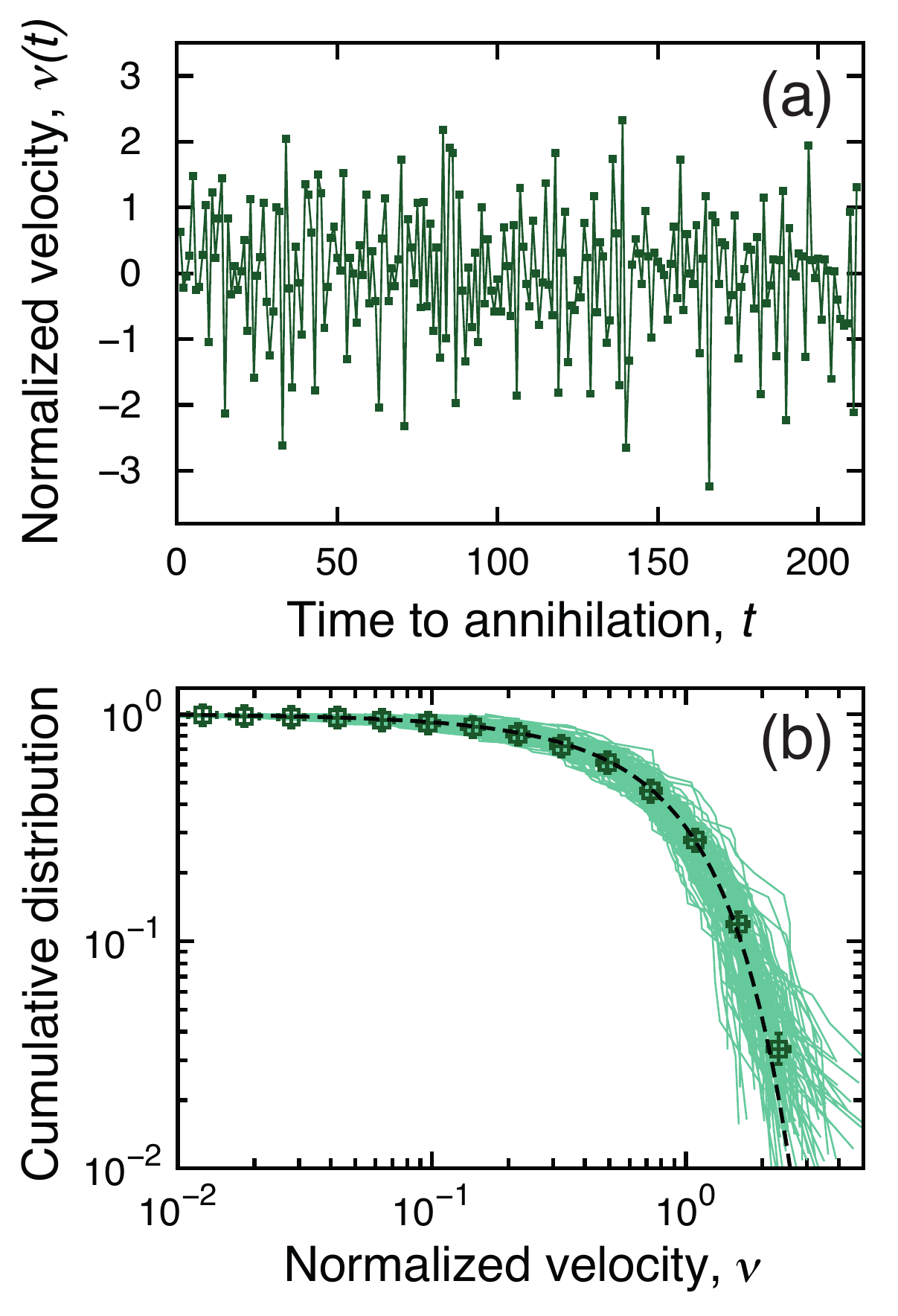}
\caption{(Color online) Statistics of the relative velocity between pairs of defects in our lyotropic system. (a) Typical time series of the normalized relative velocities $\nu(t)$. (b) Cumulative distributions of $\nu(t)$ for all the 49 experimental trajectories. The continuous lines represent both the 
positive and negative (taking the absolute values) tails, while the squares are average values over all distributions. The error bars are 95\% confidence intervals obtained via bootstrapping~\cite{bootstrap}. We note a good collapse of the empirical distributions and also that these distributions are quite similar to 
the Gaussian distribution of zero mean and unitary variance (dashed line).}
\label{fig2}
\end{figure}

We start our analysis by asking whether the probability distribution of the relative velocities follows a particular functional form. Figure~\ref{fig2}(a) shows
an example of time series of the relative velocities. Note that we have employed normalized velocities, that is, 
$\nu(t)=[V(t)-\langle V(t) \rangle]/\sigma(t)$, where $\sigma^2(t) = \langle [V(t)-\langle V(t) \rangle]^2 \rangle$ and $\langle\dots \rangle$ stands for the
average value. We observe that the normalized velocities fluctuate around zero with a variance that seems to not depend on the time $t$. We have
evaluated the cumulative distributions of these velocities for all the 49 experimental trajectories, as shown in Figure~\ref{fig2}(b). We note that
the distributions exhibit a good collapse and a profile that is quite similar to the Gaussian distribution of zero mean and unitary variance. In fact, 
the Kolmogorov-Smirnov test cannot reject the Gaussian hypothesis (on a significance level of 95\%) in $\sim80$\% of the cases. In addition, we remark that the Gaussian distribution describes almost perfectly the average values of these distributions (squares in Fig.~\ref{fig2}(b)).

\begin{figure}[!ht]
\centering
\includegraphics[scale=0.50]{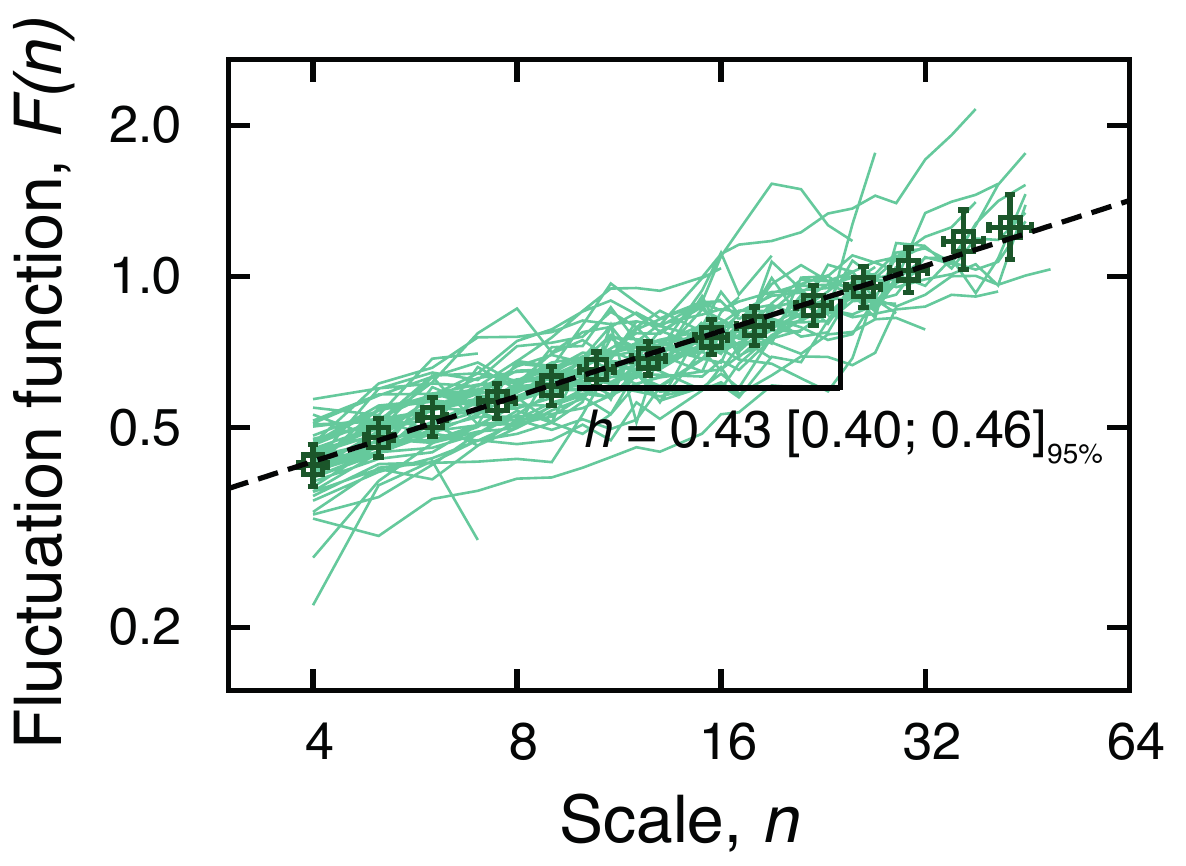}
\caption{(Color online) Long-range and anti-persistent correlations in the relative velocities in pairs of defects in our lyotropic system.
(a) Detrended fluctuation analysis (DFA) of the normalized velocities $\nu(t)$. We show the fluctuation functions $F(n)$ versus the scale $n$ for all the
49 experimental results (continuous lines) and their average values (squares). The error bars are 95\% confidence intervals obtained via 
bootstrapping~\cite{bootstrap}. We have found the mean value of the Hurst exponent to be $h = 0.43$ and the dashed line is a power law with 
this exponent. The values between brackets represent the 95\% confidence interval for $h$ obtained via bootstrapping. We remark the anti-persistent
behavior of the velocities indicated by $h<0.5$.}
\label{fig3}
\end{figure}

Another interesting question is whether there is long-range memory in the evolution of the velocities $\nu(t)$. To investigate this hypothesis, we have analyzed
the time series of the normalized velocities via detrended fluctuation analysis (DFA)~\cite{peng,bunde}. In this analysis, we first define the profile 
$Y(i) = \sum_{k=1}^{i} \nu(t_k) - \langle \nu(t_k) \rangle$ and, next, we cut $Y(i)$ into $N_n = N/n$ nonoverlapping segments of size $n$, where $N$ is the length of the series (typically $N\approx200$ in our study). For each of these segments, we fit a polynomial trend (here a polynomial of degree 1), which is subtracted from $Y(i)$, defining $Y_n(i) = Y(i) - p_{j}(i)$, where $p_{j}(i)$ is the local trend in the $j$-th segment. We thus evaluate the root-mean-square fluctuation function $F(n) = [\frac{1}{N_n} \sum_{j=1}^{N_n} \langle Y_{n}(i)^2 \rangle_j]^{1/2}$, where $\langle Y_n(i)^2 \rangle_j$ is the mean-square value of $Y_n(i)$ over the data in the $j$-th segment. For self-similar time series, $F(n)$ should display a power law dependence on the time scale $n$, that is, $F(n) \sim n^h$, where $h$ is the Hurst exponent. In our case, the DFA is shown in Fig.~\ref{fig3}. We have found that the average value of the Hurst exponents over all experimental results is $0.43$, where
the lower and upper bounds of the 95\% confidence interval are $0.40$ and $0.46$. Our results thus confirm the existence of long-range correlations in the evolution of $\nu(t)$. Moreover, since $h<0.5$, the velocities present an anti-persistent beaviour where the alternation between large and small values of $\nu(t)$ occurs  more frequently than by chance. 

\begin{figure}[!ht]
\centering
\includegraphics[scale=0.50]{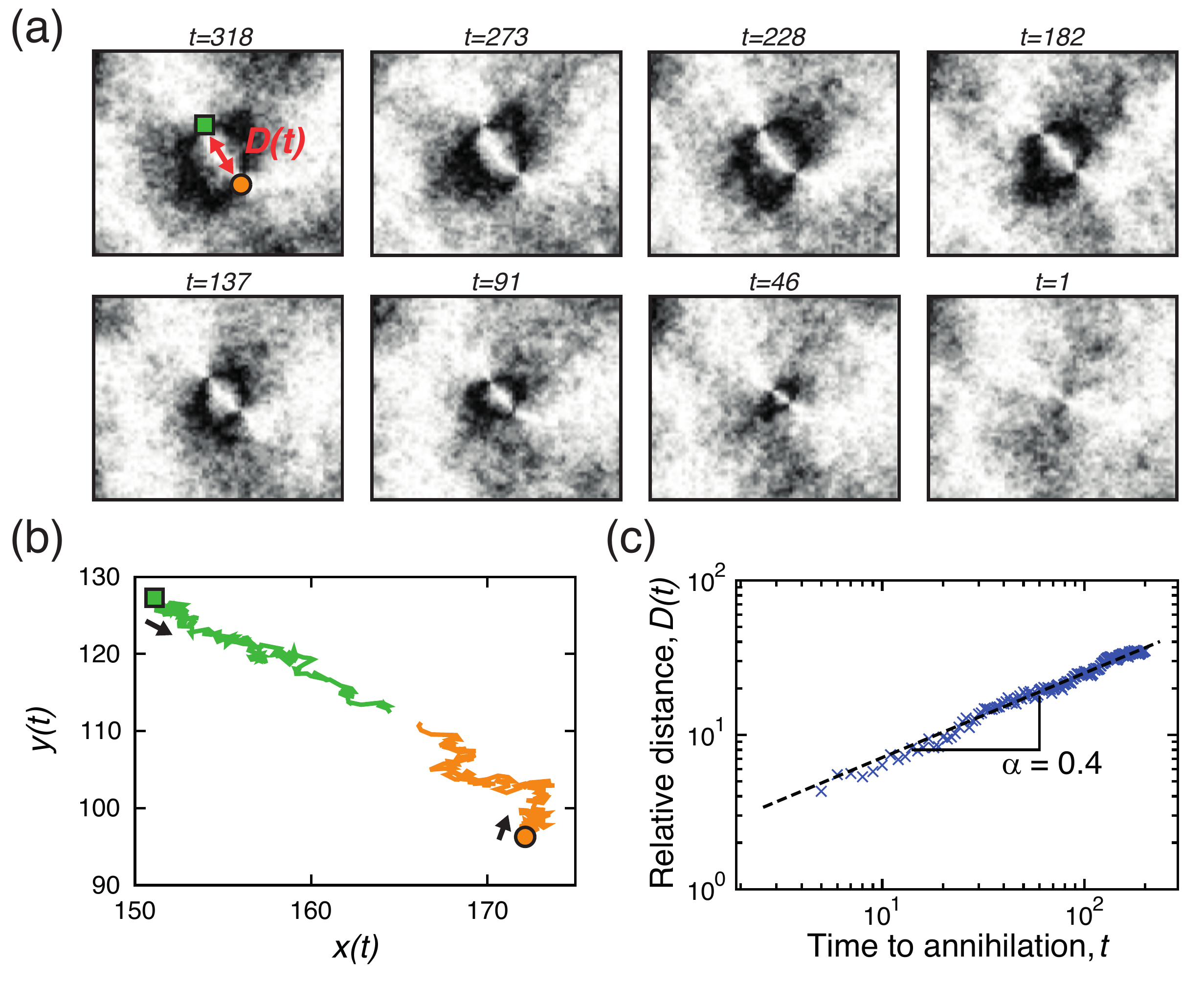}
\caption{(Color online) Annihilation process of defects obtained from the simulation of the Lebwohl-Lasher model. (a) Snapshots of a simulated texture 
of a defect pair in the course annihilation. The $t$ value above each image represents the number of iterations of the Lebwohl-Lasher model until 
the annihilation and the square and circle are the initial positions of the defects. (b) Typical trajectories of the simulated evolution of a defect 
pair obtained via the Lucas-Kanade algorithm~\cite{Lucas}. (c) The scaling law describing the relative distance between the simulated defects. 
Analogously to the experimental case, we have found $D(t) \propto t^{0.4}$.
}
\label{fig4}
\end{figure}
\textbf{\textit{Modeling -}} We now focus on modeling the previous experimental results. To this end, we have considered the well stablished Lebwohl-Lasher model~\cite{leblas,bac} using a Langevin approach~\cite{yurke}. Differently from the model $XY$ model~\cite{yurke}, which produces defects of strength $s = \pm1$, the planar Lebwohl-Lasher model showed to be able of generating topological defects with the same strength of those ones observed in our lyotropic systems, that is, $s = \pm1/2$.
This model consists of $M\times M$ spins located in a square lattice with periodic boundary conditions. The orientation of the $i$-th spin 
is given by the angle $\phi_i(\tau)$, which evolves according the Langevin-like equation~\cite{yurke}
\[
\phi_i(\tau+\Delta t)=\phi_i(\tau)-\Delta \tau \left\{ 2\,\pi\, c_{\,T} \,\eta_i + \frac{1}{2} \sum_j \,\sin[\,2(\phi_i(\tau)-\phi_j(\tau))]\right\}\,.
\]
Here, the sum over $j$ is carried out over the eight nearest neighbors of $i$, $\eta_i$ is a random number with uniform 
distribution ranging from -0.5 to 0.5, $c_{\,T}$ is a model parameter for adjusting the noise amplitude and, $\Delta t=0.05$
is the time increment. In our simulations, we have used $M=250$ and a random initial configuration, where $\phi_i(0)$ was 
chosen from a uniform distribution between $0$ and $2\pi$. This condition is analogous to the isotropic phase 
displayed by our lyotropic system at $50^\circ$ Celsius. Also, we have chosen $c_{\,T}=0.7$ because this value is
below the corresponding one of the phase transition ($c_{\,T}\approx2.6$); as pointed out previously, the temperature of $25^\circ$ Celsius is below 
the isotropic~$\to$~nematic transition of our lyotropic system. The simulations typically ran up to $\tau=5\times10^3$ and the
equivalent to lyotropic textures were obtained by building images where the gray intensity of $i$-th pixel 
is proportional to $\sin^2(2\,\phi_i)$. Figure~\ref{fig4}(a) shows snapshots of one of the 49 annihilation processes 
obtained through our simulations. 

\begin{figure}[!ht]
\centering
\includegraphics[scale=0.50]{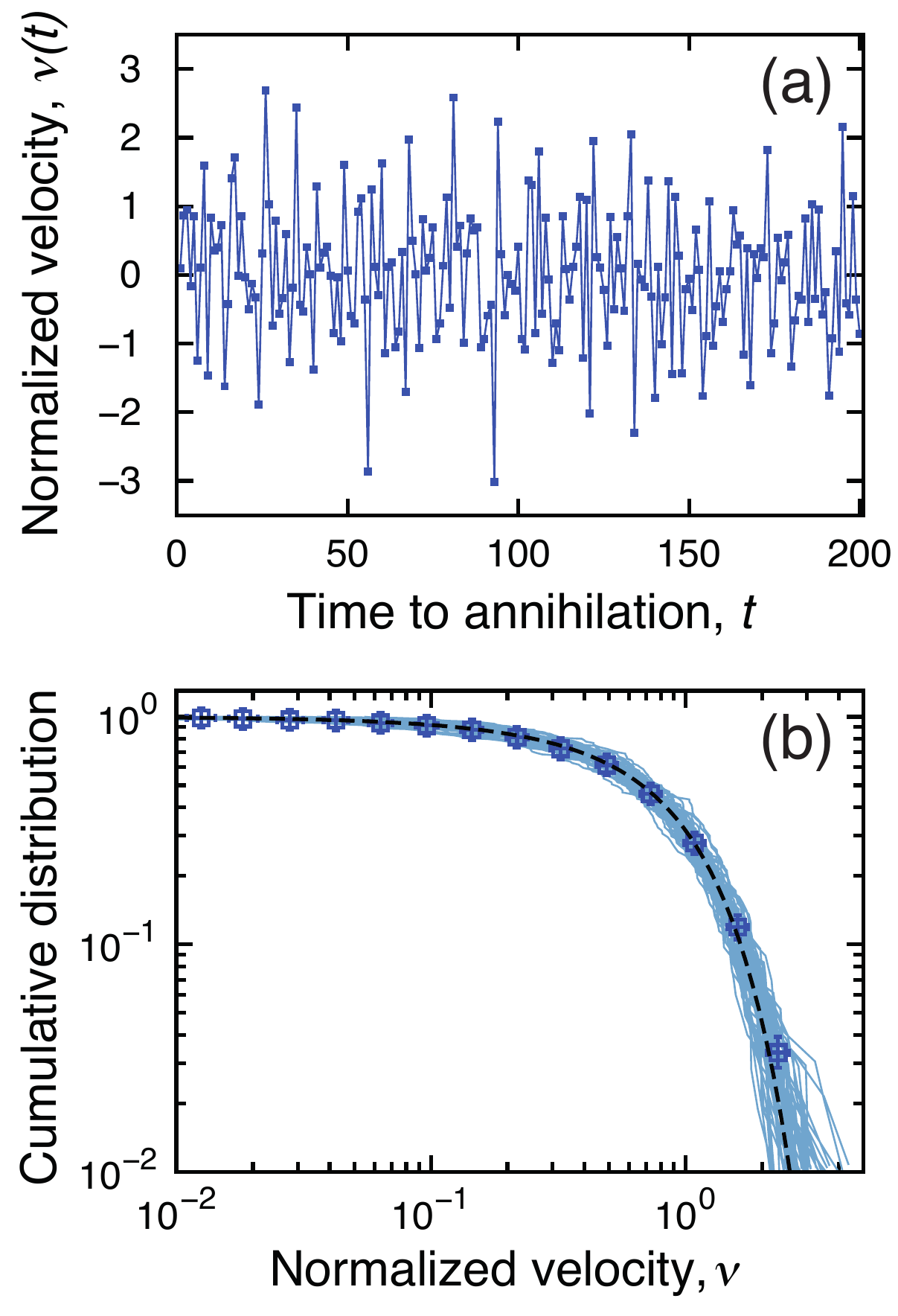}
\caption{(Color online) Statistics of the relative velocities in pairs of defects obtained through the simulation of the Lebwohl-Lasher model. 
(a) Typical time series of the normalized relative velocities $\nu(t)$. (b) Cumulative distribution of $\nu(t)$ for all the 
49 simulated trajectories. The continuous lines represent both the positive and negative (taking the absolute values) tails, 
while the squares are the averaged values over all distributions. The error bars are 95\% confidence intervals obtained via bootstrapping~\cite{bootstrap}. 
Similarly to experimental case, we have found a good collapse of the distribution and also a good agreement with the
Gaussian distribution of zero mean and unitary variance (dashed line). The the Kolmogorov-Smirnov test cannot reject the 
Gaussian hypothesis (on a significance level of 95\%) in $\sim90$\% of the cases.
}\label{fig5}
\end{figure}

\begin{figure}[!ht]
\centering
\includegraphics[scale=0.50]{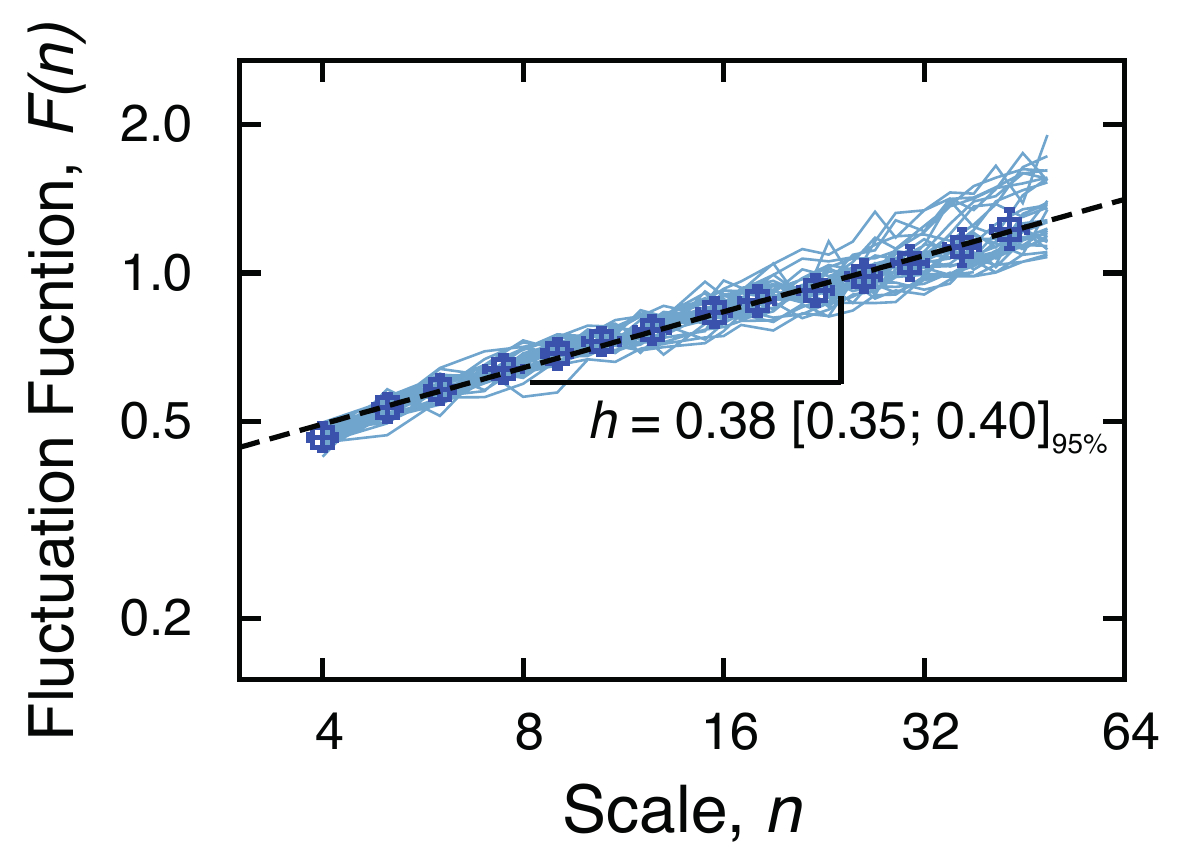}
\caption{
(Color online) Anti-persistent and long-range correlations in the relative velocities between defects obtained through the simulation of the Lebwohl-Lasher model.
(a) Detrended fluctuation analysis (DFA) of the normalized velocities $\nu(t)$. We show the fluctuation functions $F(n)$ versus the scale $n$ for all the
49 simulated results (continuous lines) and their average values (squares). The error bars are 95\% confidence intervals obtained via 
bootstrapping~\cite{bootstrap}. Similarly to experimental case, the mean value of the Hurst exponent is $h = 0.38$ and the dashed 
line is a power law with this exponent. The values between brackets represent the 95\% confidence interval for $h$ obtained via bootstrapping. 
Note the existence of overlap in the confidence intervals obtained for the experimental results and the simulated ones.
}\label{fig6}
\end{figure}

Once we have the simulated textures of the annihilation process (Fig.~\ref{fig4}(a)), we proceed as in the experimental case, that is, we applied the
Lucas-Kanade algorithm for tracking the positions of the defects (Fig.~\ref{fig4}(b)) and evaluated the relative distance 
$D(t)$. As shown in Fig.~\ref{fig4}(c), Lebwohl-Lasher model also generates a power law evolution for the distance $D(t)$ 
with the same exponent ($\alpha=0.4$) obtained for the lyotropic system (Fig.~\ref{fig1}(a)). Remarkably, the model reproduces
quite well the Gaussian distribution of the normalized velocities $\nu(t)$, as shown in Fig.~\ref{fig5}(a) and \ref{fig5}(b), and
the anti-persistent and long-range behaviors of $\nu(t)$, as shown in Fig.~\ref{fig6}. Regarding this last point, we note that
the average value of the Hurst exponent $h$ is slightly smaller in simulated case than in the experimental one. 
However, from the statistical point of view, we cannot reject the hypothesis of equality of both means since there is 
overlapping in their confidence intervals.

\textbf{\textit{Summary -}} We have thus studied the annihilation dynamics of defects in a lyotropic system. Our analysis revealed that,
in addition to the power law behavior of the relative distance $D(t)$, the motion of these defects are characterized by relative velocities $\nu(t)$
that are Gaussian distributed and present long-range correlations with an average Hurst exponent of $h=0.43$. This smaller than $1/2$ value of $h$
indicates that changes between small and large values of $\nu(t)$ occur much more frequently than by chance. In order to model the experimental
results, we have employed the Lebwohl-Lasher model due its simplicity and ability of generating the same type of topological defects present
in our lyotropic system. The results obtained from this model showed to be amazingly similar to the experimental ones. We further believe
that our work prompt new questions such as whether this anti-persistent behavior is a universal characteristic of the annihilation or it
is a system-specific behavior. 

\acknowledgements
We thank Capes, CNPq and Funda\c{c}\~ao Arauc\'aria for partial financial support. HVR is especially grateful to Funda\c{c}\~ao Arauc\'aria
for financial support under grant \mbox{number 113/2013}.

\end{document}